\documentclass[10pt,twocolumn]{article}
\usepackage[T1]{fontenc}
\usepackage[margin=0.82in]{geometry}
\usepackage{mathtools}
\usepackage{amssymb}
\usepackage{dsfont}
\usepackage{cite}
\usepackage[colorlinks=true,linkcolor=blue,citecolor=magenta,linktocpage=true]{hyperref}
\usepackage{braket}
\usepackage{titlesec}
\titleformat*{\section}{\normalsize\bfseries}
\titleformat*{\subsection}{\normalsize\bfseries}
\titleformat*{\subsubsection}{\normalsize\bfseries}

\def\be#1\ee{\begin{align}#1\end{align}}
\def\nn{\nonumber}
\def\q{\qquad}
\def\f{\frac}
\def\eps{\varepsilon}
\def\teps{\tilde{\eps}}

\def\de{\mathrm{d}}

\def\A{\mathcal{A}}

\def\E{\mathcal{E}}

\numberwithin{equation}{section}

\begin{document}

\title{\large{\textbf{\sffamily Metric formulation of the simple theory of 3d massive gravity}}}
\author{\small\sffamily Marc Geiller$^1$ \& Karim Noui$^{2,3}$}
\date{\small{\textit{
$^1$Univ Lyon, ENS de Lyon, Univ Claude Bernard Lyon 1,\\ CNRS, Laboratoire de Physique, UMR 5672, F-69342 Lyon, France\\
$^2$Institut Denis Poisson, Universit\'e de Tours, Universit\'e d'Orl\'eans, CNRS, UMR 7013, 37200 Tours, France\\
$^3$Laboratoire Astroparticule et Cosmologie, Universit\'e Paris Diderot, CNRS, UMR 7164, 75013 Paris, France\\
}}}

\twocolumn[
\begin{@twocolumnfalse}
\maketitle
\begin{abstract}
We have recently introduced a new and very simple action for three-dimensional massive gravity. This action is written in a first order formulation where the triad and the connection play a manifestly symmetric role, but where internal Lorentz gauge symmetry is broken. The absence of Lorentz invariance, which in this model is the mechanism underlying the propagation of a massive graviton, does however prevent from writing a purely metric non-linear action for the theory. Nonetheless, in this letter, we explain how to disentangle, at the non-linear level, the metric and non-metric degrees of freedom in the equations of motion. Focusing on the metric part, we show that it satisfies modified Einstein equations with higher derivative terms. As a particular case, these equations reproduce a well-studied model known as minimal massive gravity. In the general case, we obtain new metric field equations for massive gravity in three dimensions starting from the simple first order action. These field equations are consistent through a mechanism known as ``third way consistency'', which our theory therefore provides a new example of.\\\bigskip\medskip
\end{abstract}
\end{@twocolumnfalse}]

\section{Introduction}

In a recent article \cite{Geiller:2018ain}, we have introduced a new action for three-dimensional massive gravity. This action is written in the so-called first order formalism, and simply differs from the usual Hilbert--Palatini action with cosmological constant by the addition of two terms (without derivatives). The presence of these extra two terms gives a seemingly symmetric role to the triad and the connection, but does however break internal Lorentz gauge invariance. As explained through the Hamiltonian analysis \cite{Geiller:2018ain}, this breaking of Lorentz invariance is the mechanism which is responsible for the propagation of a single massive degree of freedom in this theory.

Such a mechanism is of course far from being new in field theory. In electromagnetism for example, one can give a mass to the photon by adding to the Maxwell action a term which breaks the internal U(1) gauge symmetry. This leads to the so-called Proca action \cite{Proca:1900nv}, which describes the dynamics of a massive spin-1 field propagating in Minkowski spacetime.

The extension of this mechanism to gravity (at the non-linear level) is an old issue, which was initially thought to be intractable because of the Boulware--Deser ghost \cite{Boulware:1973my}, but finally successfully addressed by de Rham, Gabadadze, and Tolley \cite{deRham:2010kj,deRham:2010ik,Hinterbichler:2011tt}, with the proof of the absence of ghost given in \cite{Hassan:2011hr}. They have proposed a theory which propagates the five degrees of freedom of a massive (four-dimensional) graviton, but which is not invariant under diffeomorphisms since it requires external fields in order to be defined (which can in turn be made dynamical, leading to bi-metric theories). This is the price to pay in order to have a non-linear theory propagating a massive spin-2 field in four spacetime dimensions.

When spacetime is three-dimensional, the story is rather different, as one can write non-linear theories of massive gravity while retaining diffeomorphism invariance. This was initially achieved by topologically massive gravity (TMG) \cite{Deser:1982vy,Deser:1981wh,Deser:1984kw}, a third order parity-breaking theory which is obtained by adding to the Einstein--Hilbert action a Chern--Simons term for the Levi--Civita connection. This was then extended in \cite{Bergshoeff:2009hq} to a fourth order parity-invariant theory known as new massive gravity (NMG), and then in \cite{Hohm:2012vh} to general massive gravity (GMG), which interpolates between TMG and NMG. Importantly, NMG propagates two coupled massive gravitons, and in this sense cannot be seen as a fundamental theory propagating an irreducible particle. This has in turn motivated the search for the most general theory of three-dimensional massive gravity propagating a single graviton, and lead to minimal massive gravity (MMG) \cite{Bergshoeff:2014pca}. Interestingly, there cannot exist a purely metric action for MMG (at the difference with TMG and NMG), and the action is written instead in a so-called Chern--Simons-like formulation \cite{Bergshoeff:2014bia,Merbis:2014vja}, where the dynamical variables are three (in the case of MMG) Lorentz algebra-valued one-forms. Nevertheless, it is possible to recast the equations of motion in a form involving a metric only (and featuring the Einstein, Cotton, and Schouten tensors). This Chern--Simons-like formulation also allows to define theories known as generalized minimal massive gravity (GMMG) \cite{Setare:2014zea} and exotic massive gravity (EMG) \cite{Ozkan:2018cxj}. In spite of the non-existence of metric actions for MMG, GMMG, and EMG, these theories have consistent (i.e. covariantly-conserved) metric field equations thanks to a mechanism known as ``third way consistency'' \cite{Bergshoeff:2015zga}. This is simply the statement that geometrical tensors appearing in the metric field equations may be conserved by virtue of the field equations themselves.

In this letter, we derive the metric field equations underlying the simple theory of massive gravity introduced in \cite{Geiller:2018ain}. We will show that in a particular case (for a specific value of a combination of the four-coupling constants of the theory) this produces the equations of MMG, while otherwise it gives rise to new modified and third-way-consistent Einstein equations (given in \eqref{modifEinstein} below) which do not fall into the class of theories mentioned above. This therefore gives a new perspective on MMG, and shows in particular that we can interpret its massive graviton as arising from a breaking of internal Lorentz invariance in the first order formulation.

\section{The simple action for 3d massive gravity}

The simple action for three-dimensional massive gravity is written in terms of a triad $e$ and a connection $\omega$ as
\be\label{action}
S
&=m_\text{p}\int e\wedge\de\omega+\f{\lambda_0}{6}e\wedge[e\wedge e]+\f{\lambda_1}{2}\omega\wedge[e\wedge e]\nn\\
&\phantom{\ =m_\text{p}\int}+\f{\lambda_2}{2}e\wedge[\omega\wedge\omega]+\f{\lambda_3}{6}\omega\wedge[\omega\wedge\omega],
\ee
where a trace in the Lorentz algebra $\mathfrak{so}(2,1)$ is understood \cite{Geiller:2018ain}. Each coupling constant $\lambda_n$ has the dimension of a mass to the power $(2-n)$. As explained at length in \cite{Geiller:2018ain}, this theory is topological when $\lambda_0\lambda_3=\lambda_1\lambda_2$, and propagates the single degree of freedom of a massive graviton otherwise.

This action can also be written in the general Chern--Simons-like formulation introduced in \cite{Bergshoeff:2014bia}. However, at the difference with the Chern--Simons-like theories which have been studied so far (see e.g. \cite{Merbis:2014vja}), the action \eqref{action} contains only two sets of dynamical variables and a single kinetic term. This is why one can think of it as being ``simple''. Notice also that, importantly, this action is not Lorentz-invariant.

Let us now turn to the key point of this letter, which is the study and the rewriting of the equations of motion. They are given by
\begin{subequations}\label{eoms}
\be
\de\omega+\f{\lambda_0}{2}[e\wedge e]+\lambda_1[\omega\wedge e]+\f{\lambda_2}{2}[\omega\wedge\omega]&=0,\\
\de e+\f{\lambda_1}{2}[e\wedge e]+\lambda_2[e\wedge\omega]+\f{\lambda_3}{2}[\omega\wedge\omega]&=0.
\ee
\end{subequations}
One can clearly see in these equations and in the action the symmetric role played by the variables $e$ and $\omega$. Indeed, one could declare that $e$ transforms as a Lorentz connection and $\omega$ as a tensor under internal gauge transformations (which in any case are \textit{not} symmetries of this theory), or the other way around, without affecting the physics.

However, in order to write down metric field equations, one would like to start by unambiguously identifying a triad variable (from which the metric is then constructed). In order to force a connection-triad interpretation upon this theory, one can look for linear combinations
\be\nn
E\coloneqq ae+b\omega,\q A\coloneqq ce+d\omega,
\ee
where $(a,b,c,d)$ are constant, such that $A$ is the Levi--Civita connection associated with $E$. One can easily show that this is indeed possible if the ratio $z\coloneqq b/a$ satisfies the equation
\be\label{zeq}
\lambda_3-\lambda_2z-\lambda_1z^2+\lambda_0z^3=0,
\ee
which always admits at least one real solution $z(\lambda_0,\lambda_1,\lambda_2,\lambda_3)$. If $a\neq0$ (which we will assume from now on), we can fix its value to $a=1$ without loss of generality. In this case, taking
\be\nn
c=\f{1}{2}(\lambda_1+z\lambda_0),\q d=\f{1}{2}(2\lambda_2+\lambda_1z-\lambda_0z^2)
\ee
enables to rewrite the equations of motion \eqref{eoms} in the desired form, i.e.
\begin{subequations}
\be
&\de E+[A\wedge E]=0,\label{eomE}\\
&\de A+\f{\gamma_1}{2}[A\wedge A]+\f{\gamma_2}{2}[E\wedge E]+\gamma_3[A\wedge E]=0,\label{eomA}
\ee
\end{subequations}
where the new coefficients are given by
\be
&\phantom{4}(\lambda_2-\lambda_0z^2)\gamma_1\coloneqq\lambda_2-2\lambda_1z+\lambda_0 z^2,\label{gamma1}\\
&4(\lambda_2-\lambda_0z^2)\gamma_2\coloneqq4\lambda_0\lambda_2^2-3\lambda_2\lambda_1^2+2(\lambda_0\lambda_1\lambda_2-\lambda_1^3)z\nn\\
&\phantom{4(\lambda_2-\lambda_0z^2)\gamma_2\coloneqq}+(\lambda_0\lambda_1^2-3\lambda_0^2\lambda_2)z^2+\lambda_0^3z^4,\nn\\
&\phantom{4}(\lambda_2-\lambda_0z^2)\gamma_3\coloneqq(\lambda_1-\lambda_0z)(\lambda_2+\lambda_1z).\nn
\ee
This of course requires that $\lambda_2-\lambda_0z^2\neq0$, which by virtue of \eqref{zeq} is always the case if the massive condition $\lambda_0\lambda_3\neq\lambda_1\lambda_2$ is satisfied.

As a consequence of this change of variables, we get the new equation of motion \eqref{eomE}, which shows that $A$ is the torsion-free connection compatible with $E$. This equation can therefore be solved to write $A(E)$, and substituting this solution into \eqref{eomA} then leads to
\be\label{eq for E1}
F+\f{\gamma_2}{2}[E\wedge E]+\f{\gamma_1-1}{2}[A\wedge A]+\gamma_3[A\wedge E]=0,
\ee
where now $F=\de A+[A\wedge A]/2$ is the curvature of the connection $A(E)$. The first two terms describe usual three-dimensional gravity with a cosmological constant, while the last two terms, which are not Lorentz-invariant, are responsible for the appearance of the propagating massive graviton. These are the field equations which we are now going to study and from which we are going to extract the modified Einstein equations for the metric only.

\section{Modified Einstein equations}

After integrating out the connection variable by solving \eqref{eomE} and writing $A=A(E)$, the dynamics of the theory is governed by the nine equations 
\eqref{eq for E1} for the remaining nine components\footnote{Here and in what follows $\mu,\nu,\rho,\dots$ are spacetime indices, and $i,j,k,\dots$ are internal $\mathfrak{so}(2,1)$ indices.} $E_\mu^i$ of the variable $E$. In principle, these components can be separated into two sets: six of them are the components of the metric $g_{\mu\nu}=E_\mu^iE_\nu^j\eta_{ij}$ and the three others are the extra non-metric components which are not gauge-invariant. 
In the equations of motion \eqref{eq for E1}, the dynamics mixes these two sets of components in a very non-trivial way. However, we are going to show how to extract from \eqref{eq for E1} equations for the metric components only. As we are going to see, these are new modified Einstein equations for massive gravity, with higher order terms. We study separately the cases $\gamma_1=1$ and $\gamma_1\neq1$, which correspond to the cases when the equations of motion \eqref{eq for E1} are respectively linear and quadratic in $A(E)$.

\subsection{Case $\boldsymbol{\gamma_1=1}$: a MMG theory}

In the case $\gamma_1=1$, we get from \eqref{gamma1} the condition that $z(\lambda_1-\lambda_0z)=0$. This equation admits two solutions. First, if $\lambda_1=\lambda_0z$, we get from \eqref{zeq} that $\lambda_3=\lambda_2z$, which corresponds to the topological sector $\lambda_0\lambda_3=\lambda_1\lambda_2$ with no massive graviton. Therefore we want to focus on the other solution, $z=0$, which from \eqref{zeq} then implies that $\lambda_3=0$. In this case the equations of motion \eqref{eq for E1} become
\be\label{eq for E2}
F-\lambda[E\wedge E]+\lambda_1[A\wedge E]=0,
\ee
where we have introduced the cosmological constant
\be\nn
\lambda\coloneqq\f{1}{2}\left(\f{3}{4}\lambda_1^2-\lambda_0\lambda_2\right),
\ee
in agreement with the analysis carried out in \cite{Geiller:2018ain}, and where once again $A=A(E)$ is the Levi--Civita connection compatible with $E$.

The non-metric degrees of freedom are hidden in the term proportional to $\lambda_1$ in equation \eqref{eq for E2}. To get rid of these non-metric degrees of freedom and obtain an equation for the metric only, we proceed in three steps.

First, we isolate the Levi--Civita connection $A$ in \eqref{eq for E2} by using the fact that the equation $[ A \wedge E]=W$  
(for any Lie-algebra valued two-form $W$) is equivalent to
\be\label{solvingA}
A_\mu^i={\eps^i}_{jk}W_{\mu\nu}^j\hat{E}^{\nu k}+\f{1}{4}E_\mu^i{\eps_j}^{kl}W_{\nu\rho}^j\hat{E}^\nu_k\hat{E}^\rho_l,
\ee
provided that the inverse $\hat{E}$ of $E$ exists. Using the fact that here $W=\lambda_1^{-1}(\lambda[E\wedge E]-F)$, we obtain that \eqref{eq for E2} is identically equivalent to an equation for the triad $E$ of the form
\be\label{eom A-B}
\E(E)\coloneqq A(E)-B(E)=0,
\ee
where $B(E)$ can be written as
\be\nn
B_\mu^i\coloneqq-\f{1}{\lambda_1}B_{\mu\nu}\hat{E}^{\nu i},\q B_{\mu\nu}\coloneqq S_{\mu\nu}-\lambda g_{\mu\nu}.
\ee
Here $S_{\mu\nu}\coloneqq R_{\mu\nu}-Rg_{\mu\nu}/4$ is the three-dimensional Schouten tensor expressed in terms of the Ricci tensor $R_{\mu\nu}$ and the Ricci scalar $R$.

The second step consists in extracting directly from the full set of equations $\E=0$ \eqref{eom A-B} those involving the metric components only. In order to achieve this, one can notice that the rewriting \eqref{eom A-B} of the equation of motion \eqref{eq for E2} is an equation for a connection, and as such is not gauge-invariant. This absence of gauge invariance in \eqref{eom A-B} is of course the same as the absence of gauge invariance in \eqref{eq for E2}. However, we now have a very natural way of transforming the equation \eqref{eom A-B} for the connection into a tensorial equation, namely by computing its curvature. We are therefore led to considering the quantity
\be\label{curvature eom}
\de\E+\f{1}{2}[\E\wedge\E]=0,
\ee
which is again trivially vanishing since it is built out of the equations of motion.

The third and final step consists in rewriting \eqref{curvature eom} in a way which depends explicitly only on the metric. After some lengthy manipulations, one obtains the following six equations for the metric:
\be\nn
\lambda_1^2\eps^{\alpha\beta\rho}R_{\mu\nu\alpha\beta}+4\lambda_1\nabla_{[\mu}{S_{\nu]}}^\rho+2\eps^{\alpha\beta\rho}B_{\mu\alpha}B_{\nu\beta}=0,
\ee
were $R_{\mu\nu\rho\sigma}$ is the Riemann tensor, $\nabla_\mu$ the covariant derivative,  $\eps$ the anti-symmetric tensor (not the symbol), and $[\mu\nu]$ denotes anti-symmetrization of indices (with weight $1/2$). One can then contract these equations with the anti-symmetric tensor $\eps^{\mu\nu\sigma}$, and use the fact that $\eps^{\mu\nu\sigma}\eps^{\alpha\beta\rho}R_{\mu\nu\alpha\beta}=4G^{\rho\sigma}$, where $G_{\mu\nu}\coloneqq R_{\mu\nu}-Rg_{\mu\nu}/2$ is the Einstein tensor, to obtain the modified Einstein equations
\be\nn
\lambda_1^2G_{\mu\nu}+\lambda_1C_{\mu\nu}+\f{1}{2}{\eps_\mu}^{\alpha\beta}{\eps_\nu}^{\rho\sigma}B_{\alpha\rho}B_{\beta\sigma}=0,
\ee
which upon expanding the last term are equivalent to
\be\label{metriceq}
(\lambda_1^2-\lambda)G_{\mu\nu}+\lambda_1C_{\mu\nu}-\lambda^2g_{\mu\nu}+J_{\mu\nu}=0.
\ee
Here  $C_{\mu\nu}\coloneqq{\eps_\mu}^{\rho\sigma}\nabla_\rho S_{\sigma\nu}$ is the Cotton tensor, and following \cite{Bergshoeff:2014pca} we have introduced $J_{\mu\nu}\coloneqq{\eps_\mu}^{\alpha\beta}{\eps_\nu}^{\rho\sigma}S_{\alpha\rho}S_{\beta\sigma}/2$. One can now finally recognize that \eqref{metriceq} are the field equations of MMG given in \cite{Bergshoeff:2014pca}, with the coupling constants there mapped to $\lambda$ and $\lambda_1$ here. In other words, the simple theory \eqref{action} with $\lambda_3=0$ is equivalent to a theory of MMG\footnote{Note that in \eqref{metriceq} it is possible to further constrain the parameters $\lambda_{0,1,2}$ (i.e. by setting some of them to zero for example), as long as one preserves the massive condition $\lambda_0\lambda_3=\lambda_1\lambda_2$.}

However, we can now appreciate a crucial difference between the action \eqref{action} and the Chern--Simons-like formulation of MMG \cite{Bergshoeff:2014bia}: this latter uses three fields and is Lorentz-invariant, while \eqref{action} uses two fields only and is not Lorentz-invariant. Furthermore, \eqref{action} reduces to MMG only when $\lambda_3=0$, and in the general case produces new modified Einstein equations, as we will show in the following section.

Finally, we can investigate the fate of the three (non-metric) degrees of freedom which are contained in the initial field equations \eqref{eq for E2} but not in the six equations \eqref{metriceq}. It turns out that these extra degrees of freedom can in fact be determined completely from the metric itself, and in this sense are not independent. Indeed, after solving \eqref{metriceq} for the metric $g_{\mu\nu}$, one can choose a corresponding triad $\bar{E}$. This triad is of course determined only up to a Lorentz transformation $u$ acting in the fundamental representation as $E=u^{-1}\bar{E}u$. This group element $u$ contains precisely the non-metric part of the triad. In order to access it, one can plug $E=u^{-1}\bar{E}u$ in the equations of motion \eqref{eq for E2} to obtain
\be\nn
F(\bar{A})-\lambda[\bar{E}\wedge\bar{E}]+\lambda_1[\bar{A}\wedge\bar{E}]+\lambda_1[U\wedge\bar{E}]=0,
\ee
where $\bar{A}\coloneqq A(\bar{E})$ and $U\coloneqq u^{-1}\de u$ is the Lorentz flat connection associated to $u$. Then, using \eqref{solvingA} allows to express $U$ and finally $u$ in terms of $\bar{E}$ only. Therefore, as announced, the extra three variables are determined by the metric itself.

\subsection{Case $\boldsymbol{\gamma_1\neq1}$: a new massive gravity theory}

The case $\gamma_1\neq1$ is much more interesting because it leads to new modified Einstein equations for three-dimensional massive gravity.

In equation \eqref{eq for E1}, we now have the last two terms which are not Lorentz-invariant, and which feature a quadratic part in the Levi--Civita connection $A(E)$. As in the previous subsection, in order to eliminate these terms and to obtain an equation for the metric we proceed in three steps.

We start by isolating the connection. For this we first introduce the new variable
\be\nn
\A\coloneqq A+\xi_3E,\q\xi_3\coloneqq\f{\gamma_3}{\gamma_1-1}=-\f{\lambda_2+\lambda_1z}{2z},
\ee
such that \eqref{eq for E1} becomes
\be\label{calAeq}
F+\f{\xi_2}{2}[E\wedge E]+\f{\xi_1}{2}[\A\wedge\A]=0,
\ee
with the new coefficients $\xi$ given by
\be
&\xi_1\coloneqq\gamma_1-1=\f{2z(\lambda_0z-\lambda_1)}{\lambda_2-\lambda_0z^2},\nn\\
&\xi_2\coloneqq\gamma_2-\f{\gamma_3^2}{\gamma_1-1}=\f{(\lambda_1+\lambda_0z)(2\lambda_2+\lambda_1z-\lambda_0z^2)}{4z}.\nn
\ee
Then we write the components of \eqref{calAeq} explicitly as
\be\nn
\xi_1\eps_{ijk}\teps^{\mu\nu\rho}\A_\nu^j\A_\rho^k=W^\mu_i\coloneqq-\teps^{\mu\nu\rho}(F_{i\nu\rho}+\xi_2\eps_{ijk}E_\nu^jE_\rho^k),
\ee
where $\teps$ is now the anti-symmetric symbol, which is in turn equivalent to the equation
\be\nn
\A_\mu^i=-2\xi_1(\det\A)\hat{W}_\mu^i,
\ee
where $\hat{W}$ is the inverse of  $W$. From this, we finally get that \eqref{calAeq} is equivalent to the equations
\be\nn
\E(E)\coloneqq\A(E)+\epsilon\hat{W}(E)\left(-\f{\det W(E)}{2\xi_1}\right)^{1/2}= 0,
\ee
which generalize \eqref{eom A-B}, and where $\epsilon = \pm 1$ is a sign inherited from the fact that \eqref{calAeq} is quadratic in $\A$. 

The second step consists in isolating the equations involving the metric only. Again, this can be done by considering the curvature \eqref{curvature eom} of the form of the equations of motion which we have just obtained.

The last step is then once again to massage this curvature equation until it takes a simple enough form. After a long manipulation, we arrive at new modified Einstein equations for three-dimensional massive gravity reading
\be\label{modifEinstein}
\left(\xi_1-1\right)G_{\mu\nu}+\left(\xi_2-\xi_1\xi_3^2\right)g_{\mu\nu}+L_{\mu\nu}=0,
\ee
where we have introduced the tensors
\be
L_{\mu\nu}&\coloneqq-\xi_1\eps_{\mu\alpha\beta}\nabla^\alpha\A^\beta_\nu-\xi_1\xi_3(\A_{\mu\nu}-\A g_{\mu\nu}),\label{def L}\\
\A_{\mu\nu}&\coloneqq\A_\mu^iE_{\nu i}=\epsilon\left(\f{\det(H_\alpha^\beta)}{\xi_1}\right)^{1/2}H^{-1}_{\mu\nu},\nn\\
H_{\mu\nu}&\coloneqq G_{\mu\nu}-\xi_2g_{\mu\nu}.\nn
\ee
These new Einstein equations feature the Einstein tensor, a cosmological term, and the new tensor $L_{\mu\nu}$. Let us now say a word about the consistency of these equations and their solution space.

Consistency of the equations of motion, which can be checked by taking their covariant divergence, requires that $\nabla^\mu L_{\mu\nu}\stackrel{\text{\tiny{!}}}{=}0$. This condition does however not follow from the above definition \eqref{def L} of the tensor $L_{\mu\nu}$, which means as expected that these equations of motion cannot be derived from a purely metric action. Instead, consistency is achieved via the so-called ``third way'' of \cite{Bergshoeff:2015zga}, i.e. using the equations of motion themselves. Indeed, by taking the divergence of \eqref{def L} one finds that\footnote{We consider here that $\xi_3\neq0$ since $\xi_3=0$ corresponds to the topological sector.}
\be\nn
\nabla^\mu L_{\mu\nu}=-\eps_{\mu\alpha\beta}\nabla^\mu\nabla^\alpha\A_\nu^\beta-\xi_3(\nabla^\mu\A_{\mu\nu}-\nabla_\nu\A)\stackrel{\text{\tiny{!}}}{=}0,
\ee
and one can then show by an explicit calculation that the first term is identically vanishing, while the second one is simply the anti-symmetric part of the equations of motion (obtained by contracting \eqref{modifEinstein} with the $\eps$ tensor). Upon using the anti-symmetric part of the equations of motion, the field equations therefore have a vanishing covariant divergence. This makes these field equations consistent, and provides another example of the third way consistency. Note however that in the present case this consistency is different from the examples which have been given previously in the literature, since here it involves a non-trivial anti-symmetric contribution to the equations of motion.

This then raises the question of the role of this anti-symmetric part of the equations of motion, since a priori \eqref{modifEinstein} are nine equations for the six components of the metric. Indeed, one could therefore worry that the equations of motion over-determine the metric in an inconsistent way. However, we have just shown that the equations of motion are consistent, and in the concluding section which follows we explain that this theory does admit maximally symmetric solutions. This therefore opens the possibility of studying perturbations over these backgrounds, and and we plan to come back to this in future work. Note that such anti-symmetric contributions to equations of motion in first order theories were also present and understood in \cite{Hassan:2012wt}.

Finally, note that the MMG field equations \eqref{metriceq} cannot be obtained from the general form \eqref{modifEinstein} since the limit $\gamma_1=1$ implies that $\xi_1=0$ and is therefore degenerate. Furthermore, since the couplings $\xi_{1,2,3}$ appearing in \eqref{modifEinstein} depend on $z$, which is the real solution to \eqref{zeq}, one might wonder what happens in the case $\lambda_0=0$ (the case $\lambda_3=0$ poses no problem, since then one has $z=0$, and this is the case of MMG treated in the previous section). In fact, it is easy to see from the symmetry of \eqref{action} under the simultaneous swaps $(e\leftrightarrow\omega,\lambda_0\leftrightarrow\lambda_3,\lambda_1\leftrightarrow\lambda_2)$ that the case $\lambda_0=0$ will actually lead back to the MMG theory \eqref{metriceq}, but where now we have $E=\omega$ instead of $E=e$.

\section{Perspectives}

Now that we have extracted metric field equations from the theory \eqref{action}, it is easy to look for exact solutions (as opposed to working with the Lorentz non-invariant connection and triad field equations). In particular, we immediately see that the new theory admits maximally symmetric solutions defined by
\be\nn
G_{\mu\nu}+\Lambda g_{\mu\nu}=0,
\ee
where $\Lambda$ is an effective cosmological constant to be determined in terms of the coupling constants of the theory. Substituting this equation in \eqref{modifEinstein} tells us that $\Lambda$ is determined by the quadratic equation
\be\nn
\Lambda^2+\f{2(\xi_1\xi_3^2+\xi_1^2\xi_3^2-\xi_1\xi_2)}{\xi_1-1}\Lambda+\f{(\xi_2+\xi_1\xi_3^2)^2}{\xi_1-1}=0.
\ee
For example, flat metrics exist only if $\Lambda=0$ is a solution of this equation, which implies that $\xi_2+\xi_1\xi_3^2=0$, which can be shown to be equivalent to the condition derived in \cite{Geiller:2018ain}, providing a good consistency check.

The BTZ black hole, being locally isometric to anti-de-Sitter spacetime, is therefore also a solution of the field equations \eqref{metriceq}. It would be very interesting to study the stability of this solution under linear perturbations, and its holographic description (possibly using the framework of \cite{Grumiller:2017otl}) in view of understanding its thermodynamical properties.

Going further, one should also investigate the possible off-shell ambiguities which may distinguish \eqref{action} from already existing massive gravity models \cite{Bergshoeff:2018luo}. This is particularly important for understanding the unitarity properties of the theory and the properties of its boundary holographic dual.

Finally, let us recall that we have obtained a new theory of massive gravity in three dimensions based on the breaking of internal Lorentz invariance in the first order formalism. We have managed to integrate out the non-metric degrees of freedom and to find modified Einstein equations involving the metric tensor only. This suggests that this new mechanism to generate a massive graviton could potentially be transposed to higher dimensions.

\section*{Acknowledgements}

We would like to warmly thank Jihad Mourad for guiding us towards the computation of the modified Einstein equation in this theory, and Wout Merbis and Paul Townsend for carefully reading and providing many useful comments on a previous draft.

\bibliography{3Dmassive_biblio}
\bibliographystyle{Biblio}

\end{document}